\documentclass[dvipsnames, twocolumn]{aastex631}

\usepackage{orcidlink}

\begin{document}

\title{A deep Search for Ethylene Glycol and Glycolonitrile in the V883 Ori Protoplanetary Disk}

\author[0009-0003-6626-8122]{Abubakar M. A. Fadul}
\affiliation{Max Planck Institute for Astronomy, Königstuhl 17, D-69117 Heidelberg, Germany}
\affiliation{Faculty of Physics, University of Duisburg-Essen, Lotharstra\ss e 1, 47057 Duisburg, Germany}

\author[0000-0002-6429-9457]{Kamber R. Schwarz}
\affiliation{Max Planck Institute for Astronomy, Königstuhl 17, D-69117 Heidelberg, Germany}

\author[0000-0002-4755-4719]{Tushar Suhasaria}
\affiliation{Max Planck Institute for Astronomy, Königstuhl 17, D-69117 Heidelberg, Germany}

  \author[0000-0002-0150-0125]{Jenny K. Calahan}
 \affiliation{Center for Astrophysics | Harvard \& Smithsonian, 60 Garden St., Cambridge, MA 02138, USA}

 \author[0000-0001-6947-6072]{Jane Huang}
 \affiliation{Department of Astronomy, Columbia University, 538 W. 120th Street, Pupin Hall, New York, NY 10027, USA}

 \author[0000-0002-2555-9869]{Merel L. R. van ’t Hoff}
 \affiliation{Department of Physics and Astronomy, Purdue University, 525 Northwestern Ave, West Lafayette, IN 47907, USA}
 
\begin{abstract}

Ethylene glycol ($\mathrm{(CH_2OH)_2}$, hereafter EG) and Glycolonitrile ($\mathrm{HOCH_2CN}$, hereafter GN) are considered molecular precursors of nucleic acids. EG is a sugar alcohol and the reduced form of Glycolaldehyde ($\mathrm{CH_2(OH)CHO}$, hereafter GA). GN is considered a key precursor of adenine formation (nucleotide) and can be a precursor of glycine (amino acid). 
Detections of such prebiotic molecules in the interstellar medium are increasingly common. How much of this complexity endures to the planet formation stage, and thus is already present when planets form, remains largely unknown. Here we report Atacama Large Millimeter/sub-millimeter Array (ALMA) observations in which we tentatively detect EG and GN in the protoplanetary disk around the outbursting protostar V883 Ori. The observed EG emission is best reproduced by a column density of $\mathrm{3.63^{+0.11}_{-0.12} \times 10^{16} \; cm^{-2}}$ and a temperature of at least 300 K. The observed GN emission is best reproduced by a column density of $\mathrm{3.37^{+0.09}_{-0.09} \times 10^{16} \; cm^{-2}}$ and a temperature of $88^{+1.2}_{-1.2}$ K. 
Comparing the abundance of EG and GN relative to methanol in V883 Ori with other objects, V883 Ori falls between hot cores and comets in terms of increasing complexity. This suggests that the build up of prebiotic molecules continues past the hot core phase into the epoch of planet formation. Nascent planets in such environments may inherit essential building blocks for life, enhancing their potential habitability. Further observations of this protoplanetary disk at higher spectral resolution are required to resolve blended lines and to confirm these tentative detection.

\end{abstract}



\section{Introduction} 
\label{sec:intro}

The ingredients for life originated in space. A key element to assessing how common life is in the universe is understanding the extent to which the Earth inherited biotic molecules. 
Within our own solar system sugars, sugar acids, amino acids, and the nucleobases which make up DNA and RNA have been seen in asteroids and meteorites \citep{Cooper01,Glavin10,Callahan11,Parker23,Oba23,Koga24,Glavin25}. 
Comets, which are comprised of less processed material than meteors and asteroids, have not been observed to contain the most complex molecules listed above. However, sugar acids thought to be the precursors of amino acids and nucleic acids, such as GA and EG, have been observed in the comae of several comets \citep{Crovisier04,LeRoy15,Biver2015SciA,Biver24}. Additionally, glycine, the simplest amino acid, which is thought to form from the reaction between GN and ammonia (\citet{Zeng19} and references therein), was detected in the coma of comet 67P/Churyumov-Gerasimenko (67P/C-G) by the Rosetta mission \citep{Altwegg16}.

Observations of hot dense cores in the ISM have revealed a plethora of complex organic molecules (COMs), molecules six atoms or larger containing at least one Carbon atom \citep{McGuire22}. This includes GA, EG, and the nitrogen-bearing GN (also called hydroxyacetonitile) among others \citep{Hollis00,Hollis02,Rivilla17,Rivilla22}.
This chemical complexity appears to persist in the envelopes of protostars \citep{Coutens2015A&A,Jorgensen2016A&A,Zeng19,Ligterink2021A&A}. 

How much this chemical complexity is preserved, enhanced, or destroyed in the protoplanetary disk remains uncertain. Shocks and heating during protostellar collapse, and as infalling gas transitions from the envelope to the protoplanetary disk, may destroy many COMs. In this `reset' scenario the COMs seen in comets must form no earlier than the protoplanetary disk phase. Even if some chemical complexity is inherited, additional ice chemistry on dust grains within the disk can lead to increased COM abundances. 

Relatively few COMs have been detected in protoplanetary disks compared to less evolved sources \citep{Walsh16,Yamato2024AJ}. The difficulty of observing COMs in protoplanetary disks is two fold. First, the emission area is small. Second, disks are cold, such that many COMs are frozen out on dust grains as ices throughout much of the disk. There are two special cases which have proved conducive to detecting COMs in disks. The first is dust traps, where photodesorbed ices can be seen in the gas \citep{Marel2021A&A,Booth24b}. The second is in FU Orionis objects during an outburst event, which heats the disk, liberating ices and increasing the area hot enough to emit \citep[e.g.,][]{Calahan24}.  This second approach has proved successful in detecting a plethora of molecules in the disk of V883 Ori, including COMs \citep{Merel2018ApJ, Lee_2019,Tobin23,Yamato2024AJ,Lee2025ApJ}. However, the more complex pre-biotic species detected in the envelopes of less evolved protostars remain undetected in any protoplanetary disk. 

Here we report tentative detection of EG and GN in the V883 Ori protoplanetary disk. Our observations cover a frequency range not yet explored for this object, but which has yielded tentative detections of pre-biotic molecules in less evolved sources. The resulting spectrum is rich in molecular lines, as recently reported by \cite{Fadul25}. In this work we focus on the detection of the pre-biotic species. Section~\ref{sec:reduction} describes the observations and data reduction. Section~\ref{sec:analysis} describes the data analysis. Section~\ref{sec:discussion} discusses our findings in the context of other sources. Finally, our conclusions are given in Section~\ref{sec:conclusion}.


\section{Observations and Data Reduction} \label{sec:reduction}

V883 Ori was observed with the Atacama Large Millimeter/sub-milliter Array (ALMA) over six execution blocks between 30 December, 2021 and 19 September, 2022 as part of project 2021.1.00452.S. Observations were done in Band 7 in spectral scanning mode, covering the frequency range 348-366 GHz. The spectral resolution was set to 488.281 kHz, corresponding to approximately 0.42 km $\mathrm{s^{-1}}$ at these frequencies.

Data reduction was carried out using the standard ALMA calibration pipeline in the Common Astronomy Software Applications (CASA) package, version 6.4.1.12. Imaging was carried out using \textit{tclean} with Briggs weighting and a robust factor of 0.5. The final continuum subtracted line cube has a $0.48\arcsec \times 0.29\arcsec$ beam with a position angle of $-68^{\circ}$. The RMS noise level is 5.97 mJy beam$^{-1}$. 
For additional details on the observations and calibration see \citet{Fadul25}.
\section{Analysis} \label{sec:analysis}

The identification and analysis of these lines was carried out with the spectral line fitting tool developed by \citet{Fadul25}. The spectroscopic parameters of EG and GN, as well as GA and the species identified by \citet{Fadul25} were taken from the Cologne Database for Molecular Spectroscopy (CDMS; \cite{CDMS2001A&A, CDMSEndres_2016}) using “Splatalogue\footnote{\url{https://splatalogue.online/\#/home}}” online database.
The model line emission is generated assuming local thermodynamic equilibrium (LTE) conditions using the formula
\begin{equation}
    F_{mod}(\nu) = \Omega \, \Delta V \, B(\nu, T_{ex}) \, \frac{\nu \, \tau_0}{2 \, c \, \sqrt{\text{ln}(2)}},
\end{equation}
where $\Omega = \mathrm{\pi (\frac{R}{d})^2}$ is the emitting area, $R = 0.3\arcsec$ is the emission radius, $d=400$ pc is the distance to the star, and $\Delta V = 2 \mathrm{ km\; s^{-1}}$ is the full width at half maximum (FWHM) of the line. The emitting area and FWHM are both based on previous higher spatial resolution observations of COMs in V883 Ori \citep{Lee_2019}. $B(\nu, T_{ex})$ is the Planck function for a blackbody, $\nu$ is the rest frequency of a given transition, and $\tau_0$ is optical depth at line center:
\begin{equation}
    \tau_0 = \Biggl(\frac{\sqrt{\mathbf{ln(2)}}}{4\pi\, \sqrt{\pi}}\Biggl)\, \Biggl(\frac{A_{ul}\, N_{tot}}{\Delta V}\Biggl)\, \Biggl(\frac{c}{\nu}\Biggl)^3\, \Biggl(\frac{n_l\, g_u}{g_l} - n_u\Biggl).
\end{equation}
In Equation 2 $A_{ul}$ is the Einstein A coefficient,  $N_{tot}$ is the total column density of the species, $n_u$ and $n_l$ are the number of molecules in the upper and lower energy levels assuming LTE, and $g_u$ and $g_l$ are the statistical weights for the respective levels. 

\begin{figure*}[!htbp]\ref{results}
    \centering
    \includegraphics[width=1.0\textwidth]{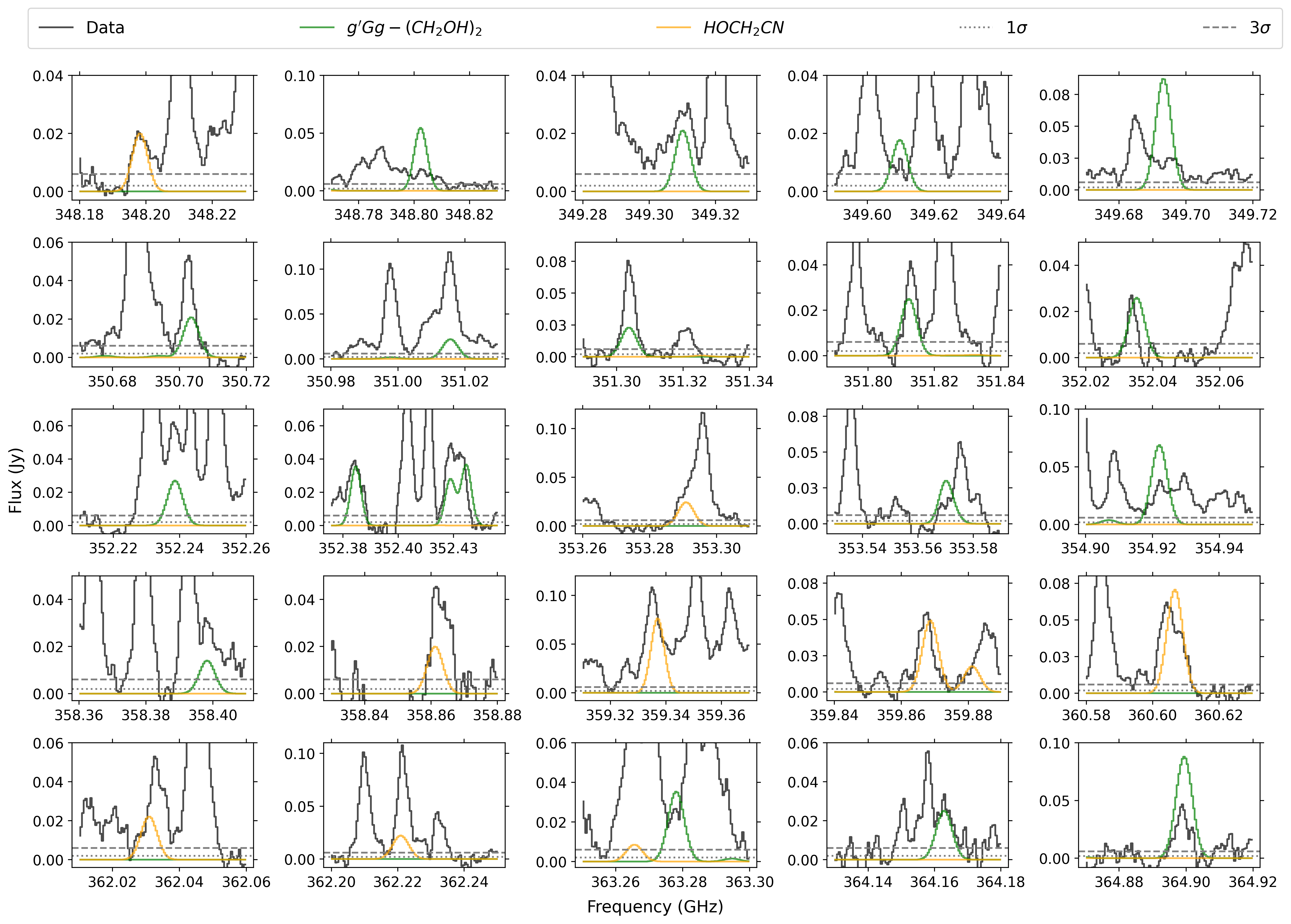}
    \caption{The best-fit model of the detected transitions of $\mathrm{g\text{'}Gg-(CH_2OH)_2}$ is shown in green, and $\mathrm{HOCH_2CN}$ is shown in orange. These are plotted on top of the data, which is shown in black, along with the summed spectra of the other detected molecules by \cite{Fadul25}, shown in blue. The dotted and dashed gray line indicates the noise level in the data (corresponding to 1 and 3$\sigma$).\label{results}}
\end{figure*}

The Markov Chain Monte Carlo (MCMC) package \textit{emcee} \citep{emcee2013} is used to identify the best fit combination of $N_{tot}$ and $T_{ex}$ for each molecule. The goodness of fit is determined by calculating the $\chi^2$ at frequencies where emission is detected above $5\sigma$ corresponding to known transitions of the fitted molecule. Fitting is done one molecule at a time, with the best fit model subtracted from the spectra before fitting subsequent species. 
The fitting and identification processes, as well as the model has been used is described in detail by \citet{Fadul25}.
We subtracted all the molecules (shown in blue in Figure \ref{results}) identified by \citet{Fadul25} from the data before fitting EG and then GN.  

We tentatively identify 15 transitions of EG and 6 transitions of GN (Figure \ref{results}). The cumulative model for all molecules detected by \cite{Fadul25}, including EG and GN, is shown in Figure~\ref{summed_spectra} in Appendix~\ref{AppendixC}. All the modeled transitions are reported in Appendix~\ref{AppendixA}. We also attempted to fit GA, based on its detection in protostars for which EG was also detected, but were unable to secure a robust detection, see Appendix D in \citet{Fadul25}.
The best fit values of $\mathrm{T_{ex}}$ and $\mathrm{N_{tot}}$ are reported in Table \ref{resulttable}. 
$\mathrm{T_{ex}}$ and $\mathrm{N_{tot}}$ are well constrain for GN by the model (see Appendix \ref{AppendixB}). 
For EG, $T_{ex}$ was not well constrained by the model, instead always approaching the highest allowed temperature. 
However, we find that using a two temperature model neither improves the fit nor provides a better constraint on the temperature(s). We fixed the temperature to 300 K where we have a best fit from the model. However, it is possible that the data would be better fit by a model with a more complex temperature structure.

\begin{figure*}[!htbp]
    \centering
    \includegraphics[width=1.0\textwidth]{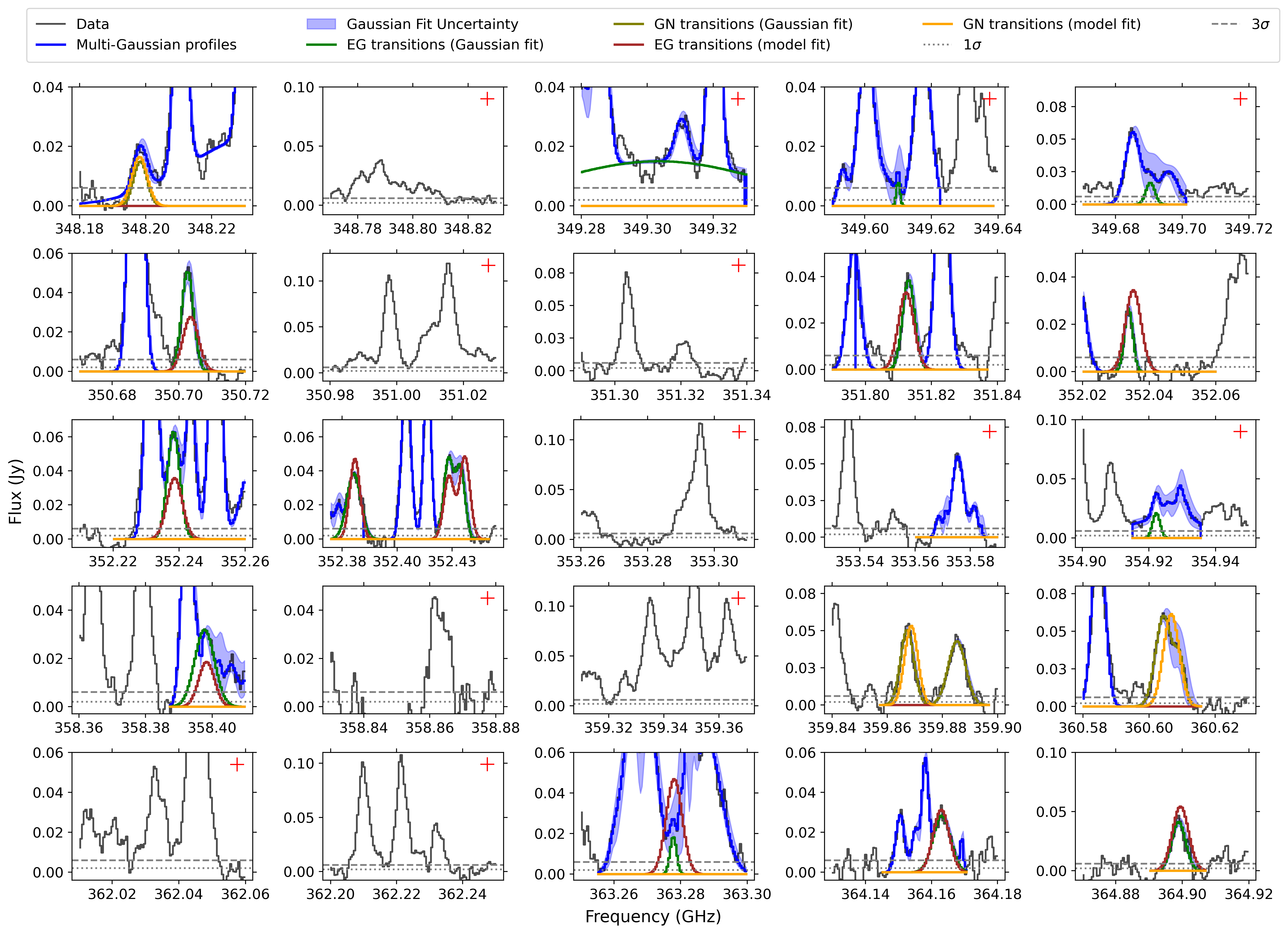}
    \caption{Demonstrates the test we performed to improve the fit. The blue line shows the Gaussian profile and its associated uncertainty. The black represents the observed data. The green and olive lines represent the Gaussian profiles of EG and GN transitions, respectively, extracted from the multi-Gaussian profile (in blue). The brown and orange lines show the best-fit of our model to the multi-Gaussian profiles of the EG and GN transitions, respectively. Panels marked with a red cross are excluded from the fit because the transitions are either strongly blended with other molecules or relatively weak. The large FWHM in the Gaussian fit to EG in the third panel is clearly inconsistent with line emission, leading to the exclusion of this panel from the mcmc fitting.} The dotted and dashed gray line indicates the noise level in the data (corresponding to 1 and 3$\sigma$).\label{fittest}
\end{figure*}

To improve the fit we utilize a multiple Gaussian fitting approach in the region where transitions from EG or GN are present, fitting Gaussian profiles to all lines which overlap with target molecule. We then isolate the Gaussian profiles associated with EG and GN from nearby lines. Subsequently, we run our MCMC model on the isolated Gaussian fits to identify the best-fit parameters. Figure~\ref{fittest} shows the isolated Gaussian profiles of the EG and GN transitions plotted on the observed data along with the best-fit model of these profiles. No significant improvement was observed in the fit compared to the best-fit model applied directly to the observed data.

Furthermore, to confirm whether these transitions originate from EG and GN or other molecules, we checked the Splatalogue database to search for molecules that emit at similar frequencies. We modeled the emission of molecules that are highly likely to blend with EG and GN, but none were found to overlap with these two species. The molecules tested include: Acetic acid, Glycine, Aminoacetonitrile, Ethanol, Propenal, Methoxymethanol, Vinyl alcohol, among others already examined by \cite{Fadul25}.

\begin{figure*}[!htbp]
    \centering
    \includegraphics[width=1.0\textwidth]{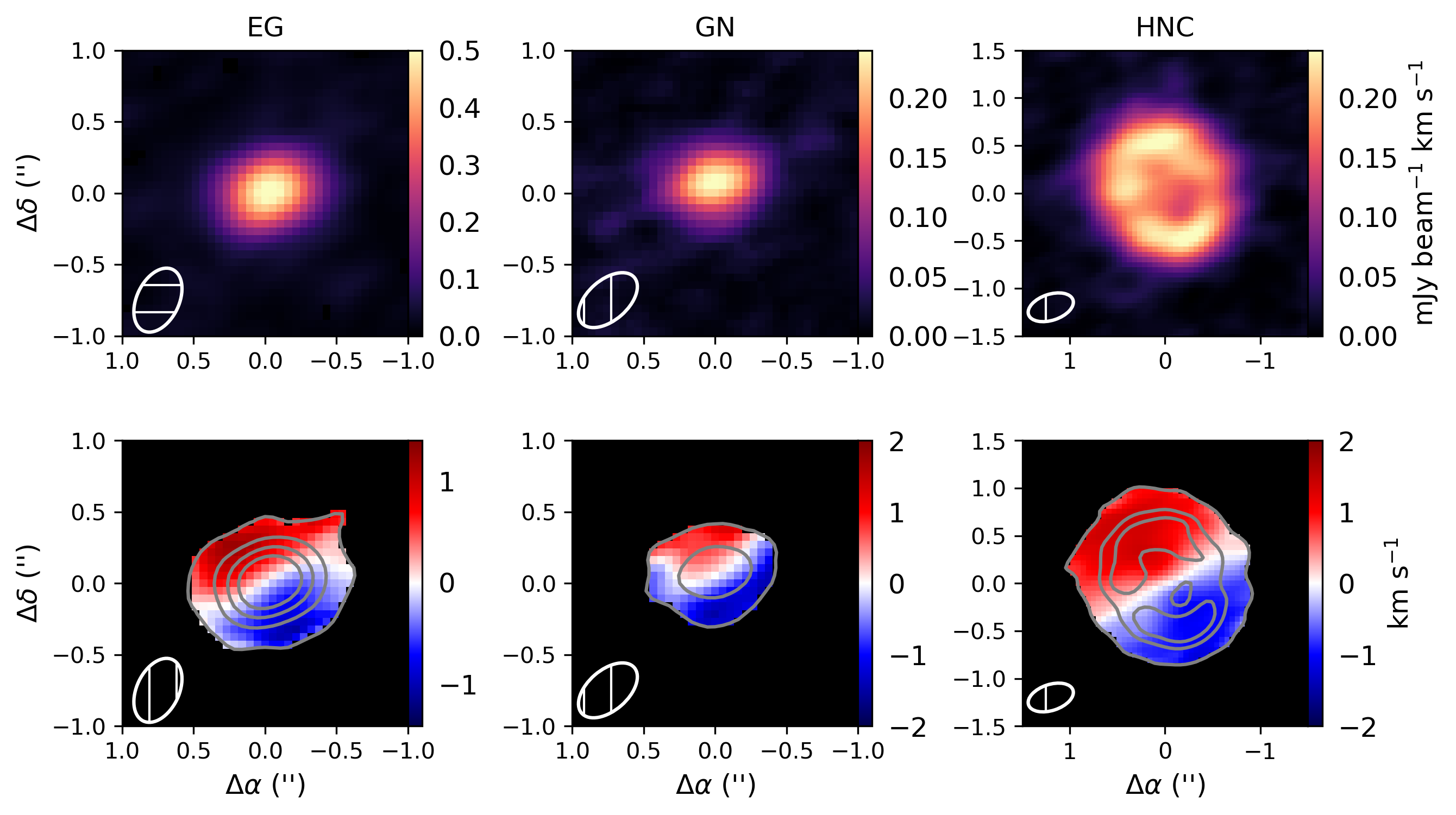}
\caption{The top and lower left panels show the integrated intensity map and the intensity-weighted velocity for EG. The top and lower right panels show the same for GN. We only included that data above $\mathrm{0.005\; Jy\, beam^{-1}}$ in the moment maps. Gray contours in the intensity-weighted velocity maps represent 1, 3, 5, and 7$\sigma$, where $\sigma = 0.005\ \mathrm{Jy; beam^{-1}; channel^{-1}}$, integrated over 10 channels. The beam size is shown in the lower left of each panel. The moment 0 maps for EG and GN were generated by summing the following transitions (EG: 350.70, 351.02, 351.30, and 351.81 GHz; GN: 348.19 and 362.22 GHz). The moment 1 maps were generated by averaging the same transitions used for moment 0 maps. The HNC detection at 362.63 GHz is shown for comparison.} \label{momentmaps}
\end{figure*}

Figure~\ref{momentmaps} shows the integrated intensity (moment 0) and the intensity weighted velocity (moment 1) maps of the most isolated EG and GN transitions.
The integrated intensity maps sum the emission from 4 transitions of EG and 2 transitions of GN respectively, while the intensity weighted velocity maps are averages. The moment 1 maps show a clear shift from red-shifted to blue-shifted across the disk semi-major axis, consistent with emission from a rotating disk. This pattern supports our assumption of emission from a Keplerian disk centered at these frequencies.


\begin{table*}[htbp]
\caption{Results of the Spectral analysis}
\centering
\begin{tabular}{lcccc}
\hline
\hline
Molecule & Formula & N$_{\text{tot}}$ (cm$^{-2}$) & T$_{\text{ex}}$ (K) & No. lines \\
\hline
Ethylene glycol & $\mathrm{g'Gg-(CH_2OH)_2}$ & $3.62^{+0.11}_{-0.12}\times 10^{16}$ & [300] & 18 \\
Glycolonitrle & $\mathrm{HOCH_2CN}$ & $3.37^{+0.09}_{-0.09}\times 10^{16}$ & $88^{+1.2}_{-1.2}$ & 9 \\
\hline
\label{resulttable}
\end{tabular}
\end{table*}

\section{Discussion}\label{sec:discussion}
\subsection{Comparison to other sources}
\begin{figure*}[!htbp]
    \centering
    \includegraphics[width=1.0\textwidth]{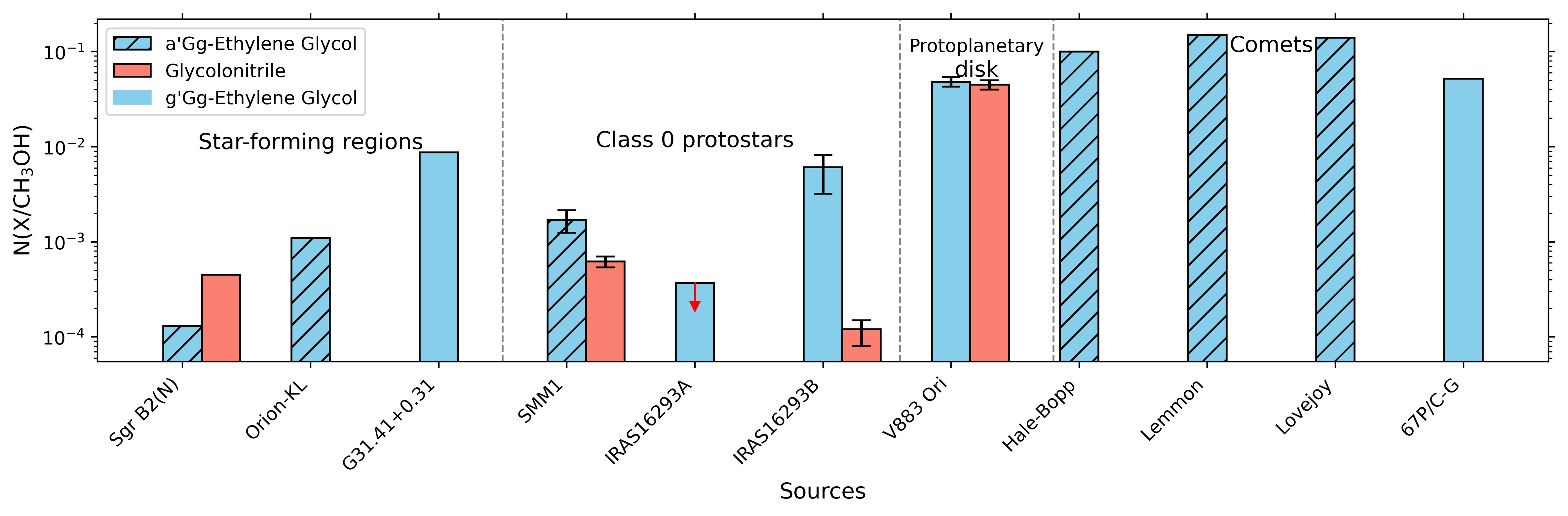}
    \caption{Comparison of EG and GN abundances relative to methanol with the high mass star-forming region Sagittarius B2 North (Sgr B2(N) a'Gg-conformer; \cite{Belloche2013A&A}), the Orion-Kleinmann Low (Orion-KL; a'Gg-conformer \cite{Brouillet2015A&A}), the molecular could G31.41+0.31 (g'Gg-conformer) \cite{Mininni2023A&A}, Class 0 Protostars: Serpens SMM1-a (a'Gg-conformer; \cite{Ligterink2021A&A}), IRAS16293-2422 A and B (g'Gg-conformer; \cite{Manigand2020A&A, Nazari2024A&A}), the protoplanetary disk V883 Ori (g'Gg-conformer; this work), and Comets: Hale-Bopp (a'Gg-conformer \cite{Bockel2000A&A}, Lemmon, and Lovejoy (a'Gg-conformer; \cite{Biver2014A&A}), and the solar system comet 67P/Churyumov-Gerasimenko \cite{Schuhmann2019ESC}. The red down arrow indicate an upper limit due to tentative detection. Bars with hatching indicate the a'Gg-conformer, while bars without hatching represent the g'Gg-conformer. The conformer of EG in comet 67P/C-G is unknown. Vertical dashed lines separate different evolutionary stages.\label{prebiotic_methanl}}
\end{figure*}

EG and GN have been detected in multiple environments, including in the central molecular zone of the Galaxy, nearby high mass star-forming regions, Class 0 protostars, and in solar system bodies. The results presented here are tentative detection of these prebiotic molecules in a protoplanetary disk. To compare the abundances across sources, we take the ratio of the derived EG and GN column density to that of methanol (Figure~\ref{prebiotic_methanl}). Methanol is one of the simplest CHO-bearing molecules, and thus one of the first to form. Thus, it is often used as a reference species when comparing abundances between sources \citep{Drozdovskaya2018MNRAS}. 

The abundance of EG relative to methanol in V883 Ori is greater than that seen in Class 0 protostellar envelopes but less than that of comets. 
GN has yet to be detected in any solar system bodies, but has been detected in a number of protostellar envelopes and molecular clouds. The abundance ratio of GN relative to methanol in V883 Ori is several orders of magnitude higher than the ratio in less evolved environments. Taken together, these ratios point to substantial prebiotic chemistry during the evolution from protostellar envelope to protoplanetary disk. 

The increased abundance of these COMs from the protostellar envelope to the protoplanetary disk stage offers insight into whether chemical complexity is lost between the two stages. 
Chemical models of protoplanetary disk environments require timescales of several million years to build up a large abundance of methanol from CO on the surface of dust grains \citep{Bosman18,Schwarz2019ApJ}. The build up of more complex species such as EG and GN should take even longer. Thus, the increase in the abundance of both EG and GN relative to methanol between envelopes and the V883 Ori disk suggests that the envelope COMs are delivered to the disk largely intact, but COMs continue to be produced in the disk. Comparison of the abundance of simpler ices in  envelopes, disks, and comets also supports the idea that chemical complexity is inherited from earlier evolutionary stages \citep{Lippi24}. 


\subsection{Upper limits of GA}\label{upper_GA}

Previous studies of protoplanetary disks have searched unsuccessfully for both GA and EG \citep{Brunken2022A&A, Lee2025ApJ}. Our observations focus on a higher frequency range in part because this regime should provide the highest signal-to-noise ratio for GA emission assuming a gas temperature of 103, the temperature derived for CH$_\mathrm{3}$OCHO by \citet{Lee_2019}. While we searched our observations for GA, we were unable to secure a robust detection. There are only two transitions that matched the data. However, these GA transitions overlap with formaldehyde ($\mathrm{H_2CO}$) transitions. For all other expected transitions there is no emission in the observed data.
Therefore, we take column density from our best fit model as an upper limit. The expected GA transitions from our model are shown in Appendix D in \cite{Fadul25}. 

The non-detection of GA could be attributed to its destruction by OH, as suggested by \cite{Coutens2018MNRAS}. Their models indicate that in low-mass sources, OH efficiently depletes GA, lowering its abundance to extremely low levels ($\sim 10^{-16}$) within a few hundred thousand years, making it undetectable. Another possible reason for its non-detection could be its conversion to EG by hydrogenation \citep{leroux2021}. These two mechanisms could significantly lower the GA abundance. Consequently, it might be extremely hard to secure a robust detection of this molecule.

\subsection{Previous non-detections of EG in V883 Ori}
Previous studies has already searched for both GA and EG \citep{Brunken2022A&A, Lee2025ApJ} in Oph IRS 48 and V883 Ori without success. For EG, we tentatively detect the g'Gg-conformer but not the a'Gg-conformer. To further test our model, we run a check using ALMA band 6 data, which published by \cite{Lee2025ApJ} as it covers many transitions of EG and GA.
We plot the predicted EG in Band 6 using our best fit results from Band 7 in Appendix~\ref{AppendixD}. While the current study does not focus on the Band 6 data, our model appears consistent with the observations in this Band. We note that the predicted EG emission is relatively weak, which could explain the non-detection reported by \cite{Lee2025ApJ}. Additionally, confusion with nearby lines could also contribute \citep[e.g.,][]{Jorgensen2020ARA&A}. In summary, the Band 6 data appear to support the existence of EG in V883 Ori, though a more detailed analysis is needed.


\subsection{Formation routes}
GN can form on the icy grain surface through several reaction pathways. First is the reaction between $\mathrm{H_2CO}$ and isocyanide (HNC) \citep{Woon2001JPCA}. The reaction is exothermic and has enough energy to desorb the formed GN into the gas phase. Second is the thermal reaction between $\mathrm{H_2CO}$, $\mathrm{NH_3}$ and HCN \citep{Danger2012ApJ}. GN can also form through the hydrogenation of $\mathrm{OCH_2CN}$ and HOCHCN \citep{Danger2013A&A}.





EG is likely formed on grain surfaces, either through the hydrogenation of GA \citep{leroux2021} or the recombination of $\mathrm{CH_2OH}$ radicals produced from methanol \citep{Butscher2015MNRAS}. Both EG and GA may originate from radical-radical reactions involving methanol. Additionally, it is known that these two molecules can interconvert via hydrogenation or irradiation processes \citep{Fedoseev2015MNRAS, hudson2005}. If EG forms from GA, the abundance ratio of the two species is expected to be near unity \citep{Rivilla17}. However, since the GA was not detected in our observations and our reported ratio of EG/GA is a lower limit, it is possible that the actual ratio being near unity cannot be ruled out. This could bring the ratio in V883 Ori into agreement with that of the Class 0 protostar IRAS 16293 B (\citet{Nazari2024A&A, Jorgensen2016A&A}). Tighter constraints on the EG/GA ratio could clarify if EG was actually formed via hydrogenation of GA.
The non-detection of GA in our observations is consistent with $\mathrm{(CH_2OH)_2} /\ \mathrm{CH_2(OH)CHO} \leq 1.6 $ for a gas temperature of 300 K. Without tighter constraints on the GA abundance, we cannot determine the formation mechanism for EG in this source. For both EG and GN, it is likely these species formed via reactions between ices on the grain surface and are observable in V883 Ori because of the outburst induced temperature increase.

\subsection{Uncertainty in the measured abundances}

The $\mathrm{CH_3OH}$ column density is taken from \cite{Fadul25}. Since $\mathrm{CH_3OH}$ column density is not reliably measured, its difficult to constrain from the main isotopologue. Instead we use the $\mathrm{^{13}CH_3OH}$ column density multiplied by the canonical ISM $\mathrm{^{12}C/^{13}C}$ ratio of 60 \citep{Langer93ApJ}. 
The ratio of $\mathrm{^{12}C/^{13}C}$ in star-forming regions has been found to be significantly lower than the canonical ISM value (see, e.g., \citealt{Jorgensen2018A&A, Yamato2024AJ, 2025arXiv250202873H}). This variation could lead to significant uncertainty, especially when using the ISM $\mathrm{^{12}C/^{13}C}$ ratio to compare the abundances of molecules in star-forming regions, where the $\mathrm{^{12}C/^{13}C}$ ratio is lower than in the ISM. Additional sources of uncertainty include dust opacity and assuming a constant source size. All of these factors have been discussed in detail in \cite{Fadul25}. Furthermore, since we tentatively detect these molecules, this also introduces uncertainty in the comparison with other sources that have confirmed detections. In particular, there is significant line blending at the current spectral resolution of 0.42 km $\mathrm{s^{-1}}$, making it difficult to disentangle contributions from other species. Additional observations of this disk are required to confirm the detection and to reliably compare it with other sources.

\section{Conclusions} \label{sec:conclusion}
We report tentative detection of 15 transitions of EG and 6 transitions of GN toward the outbursting source V883 Ori. This is the first time either species has been tentatively detected in a protoplanetary disk. We find the observations are best fit by a model of EG with a column density $3.63^{+0.11}_{-0.12}\times10^{16}$ cm$^\mathrm{-2}$, where the temperature has been fixed to 300 K. GN is best fit by a column density of $3.37^{+0.09}_{-0.09}\times10^{16}$ cm$^\mathrm{-2}$ and a temperature of $88^{+1.2}_{-1.2}$~K. Comparing these abundances to those in other sources places V883 Ori in a sequence of increasing complexity with evolutionary stage, with higher abundances than seen in protostellar cores but lower abundances than seen in comets. This further suggests that protoplantary disks inherit complex molecules from protostellar envelopes and that COM formation can continue during the protoplantary disk stage. Further observations are needed to confirm the detection of these species and further look for more complex species.


\begin{acknowledgments}
We thank the reviewer for their constructive and insightful comments, which have helped improve the quality of the paper.
This paper makes use of the following ALMA data: ADS/JAO.ALMA\#2021.1.00452.S. ALMA is a partnership of ESO (representing its member states), NSF (USA) and NINS (Japan), together with NRC (Canada) and NSC and ASIAA (Taiwan), in cooperation with the Republic
of Chile. The Joint ALMA Observatory is operated by ESO, AUI/NRAO and NAOJ. 

K.S. and T.S. would like to acknowledge European Research Council under the Horizon 2020 Framework Program via the ERC Advanced Grant Origins 83 24 28.

\end{acknowledgments}

%

\vspace{5mm}
\facilities{ALMA}


\software{Astropy \citep{2013A&A...558A..33A,2018AJ....156..123A, astropy:2022},  Gofish \citep{GoFish}, Emcee \citep{emcee2013}, CASA \citep{The_CASA_Team_2022}, Numpy \citep{numpyharris2020array}, Matplotlib \citep{matplotlibHunter:2007}, corner.py \citep{corner}, Pandas \citep{pandasmckinney-proc-scipy-2010}}

\section*{ORCID iDs}

\noindent Abubakar M. A. Fadul \orcidlink{0009-0003-6626-8122}: \url{https://orcid.org/0009-0003-6626-8122}

\noindent Kamber R. Schwarz \orcidlink{0000-0002-6429-9457}: \url{https://orcid.org/0000-0002-6429-9457}

\noindent Tushar Suhasaria \orcidlink{0000-0002-4755-4719}: \url{https://orcid.org/0000-0002-4755-4719}

\noindent Jenny K. Calahan \orcidlink{0000-0002-0150-0125}: \url{https://orcid.org/0000-0002-0150-0125}

\noindent Jane Huang \orcidlink{0000-0001-6947-6072}: \url{https://orcid.org/0000-0001-6947-6072}

\noindent Merel L. R. van ’t Hoff \orcidlink{0000-0002-2555-9869}: \url{https://orcid.org/0000-0002-2555-9869}




\bibliography{sample631}{}
\bibliographystyle{aasjournal}



\appendix\label{secA2}



\section{Appendix A}\label{AppendixA}
Appendix~\ref{AppendixA} presents a list of spectral transitions tentatively identified in our analysis.

\begin{longtable}{cccccl}
\caption{Tentative detection of EG and GN Transitions\label{tab:transitions}} \\
\hline \hline
\multicolumn{1}{c}{Transition} & \multicolumn{1}{c}{Frequency} & \multicolumn{1}{c}{Einstein A} & \multicolumn{1}{c}{$g_u$} & \multicolumn{1}{c}{Energy upper} & \multicolumn{1}{c}{Blended} \\
\multicolumn{1}{c}{} & \multicolumn{1}{c}{$\nu_0$ (GHz)} & \multicolumn{1}{c}{log$_{10}A_{ul}$ (s$^{-1}$)} & \multicolumn{1}{c}{} & \multicolumn{1}{c}{$E_u$ (K)} & \multicolumn{1}{c}{} \\
\hline
\endfirsthead

\caption*{Continuation of Table~\ref{tab:transitions}}\\
\hline \hline
\multicolumn{1}{c}{Transition} & \multicolumn{1}{c}{Frequency} & \multicolumn{1}{c}{Einstein A} & \multicolumn{1}{c}{$g_u$} & \multicolumn{1}{c}{Energy upper} & \multicolumn{1}{c}{Blended} \\
\multicolumn{1}{c}{} & \multicolumn{1}{c}{$\nu_0$ (GHz)} & \multicolumn{1}{c}{log$_{10}A_{ul}$ (s$^{-1}$)} & \multicolumn{1}{c}{} & \multicolumn{1}{c}{$E_u$ (K)} & \multicolumn{1}{c}{} \\
\hline
\endhead

\hline \hline
\endfoot

\hline
\multicolumn{6}{c}{\textbf{Ethylene Glycol $\mathrm{g'Gg-(CH_2OH)_2}$}} \\
\hline
$37_{1,36} - 36_{2,35} \; , v = 0$ & 348.802586 & -$3.42435$ & 675 & 326.06599 & No \\
$30_{6,25} - 29_{5,24} \,, \, v = 0$ & 349.3103568 & -$3.86474$ & 549 & 246.11554 & Yes \\
$35_{4,31} - 34_{5,30}  \, , \, v = 0$ & 349.69353 & -$3.20425$ & 639 & 319.51415 & No \\
$47_{18,29} - 47_{17,30} \, , \, v = 0$ & 350.703786 & -$3.6413$ & 855 & 707.73134 & No \\
$46_{18,28} - 46_{17,29} \, , \, v = 0$ & 351.0159031 & -$3.64404$ & 651 & 684.79508 & Yes \\
$45_{18,27} - 45_{17,28} \, , \, v = 0$ & 351.3039335 & -$3.6469$ & 819 & 662.34685 & Yes \\
$43_{18,25} - 43_{17,46} \, , \, v = 0$ & 351.8125551 & -$3.65334$ & 783 & 618.91426 & No \\
$42_{18,24} - 42_{17,25} \, , \, v = 0$ & 352.0354431 & -$3.657$ & 595 & 597.92987 & No \\
$41_{18,23} - 41_{17,24} \, , \, v = 0$ & 352.2388443 & -$3.66099$ & 581 & 577.43358 & No \\
$36_{3,3} - 35_{4,32} \, , \, v = 0$ & 352.381056 & -$3.48615$ & 511 & 328.35156 & No \\
$36_{4,33} - 35_{3,32} \, , \, v = 1$ & 352.4266183 & -$3.48771$ & 511 & 328.41202 & No \\
$36_{4,33} - 35_{3,32} \, , \, v = 0$ & 352.4310121 & -$3.58553$ & 657 & 328.35252 & No \\
$27_{18,9} - 27_{17,10} \, , \, v = 0$ & 353.5701978 & -$3.79924$ & 495 & 341.73379 & No \\
$33_{18,15} - 33_{17,17} \, , \, v = 1-0$ & 353.9290933 & -$3.75616$ & 469 & 431.06459 & No \\
$36_{5,31} - 35_{6,30} \, , \, v = 1$ & 358.3990899 & -$3.56208$ & 657 & 344.14887 & No \\
$23_{7,16} - 22_{6,17} \, , \, v = 0$ & 363.2782984 & -$3.65169$ & 423 & 158.99611 & No \\
$21_{8,14} - 20_{7,13} \, , \, v = 0$ & 364.1631555 & -$3.67378$ & 301 & 144.01107 & No \\
$13_{12,1} - 12_{11,2} \, , \, v = 0$ & 364.8996675 & -$3.32904$ & 189 & 114.36793 & No \\
\hline
\multicolumn{6}{c}{\textbf{Glycolonitrile $\mathrm{HOCH_2CN}$}} \\
\hline
$38_{2,37} - 37_{1,36} \, , \, v = 0$ & 348.1984298 & -$3.57192$ & 77 & 329.59594 & No \\
$40_{0,40} - 39_{1,39} \, , \, v = 0$ & 353.2911373 & -$3.4031$ & 81 & 351.06461 & Yes \\
$29_{3,27} - 28_{2,26} \, , \, v = 0$ & 358.8614609 & -$4.06515$ & 59 & 205.25471 & No \\
$17_{4,13} - 16_{3,14} \, , \, v = 0$ & 359.3372576 & -$3.81874$ & 35 & 328.35252 & Yes \\
$39_{9,31} - 38_{9,30} \, , \, v = 0$ & 359.8689093 & -$2.86629$ & 79 & 457.84286 & No \\
$39_{12,27} - 38_{12,26} \, , \, v = 0$ & 359.8813728 & -$2.88562$ & 79 & 544.98815 & No \\
$39_{4,36} - 38_{4,35} \, , \, v = 0$ & 360.6069554 & -$2.84584$ & 79 & 368.30635 & No \\
$41_{0,41} - 40_{1,40} \, , \, v = 0$ & 362.0311982 & -$3.36973$ & 83 & 368.44463 & No \\
$41_{1,41} - 40_{0,40} \, , \, v = 0$ & 362.2211371 & -$3.3691$ & 83 & 368.44871 & Yes \\
\hline 
\end{longtable}

\clearpage
\section{Appendix B}\label{AppendixB}
Posterior distribution of glycolonitrile. We ran an MCMC with 100 walkers for 10,000 steps. We discarded the first 2,000 steps as burn-in, and the remaining 8,000 steps were used to demonstrate the posterior distribution of the parameter space of $\mathrm{T_{ex}}$ and $\mathrm{N_{tot}}$.

\begin{figure}[!htbp]
\includegraphics[width=1.0\textwidth]{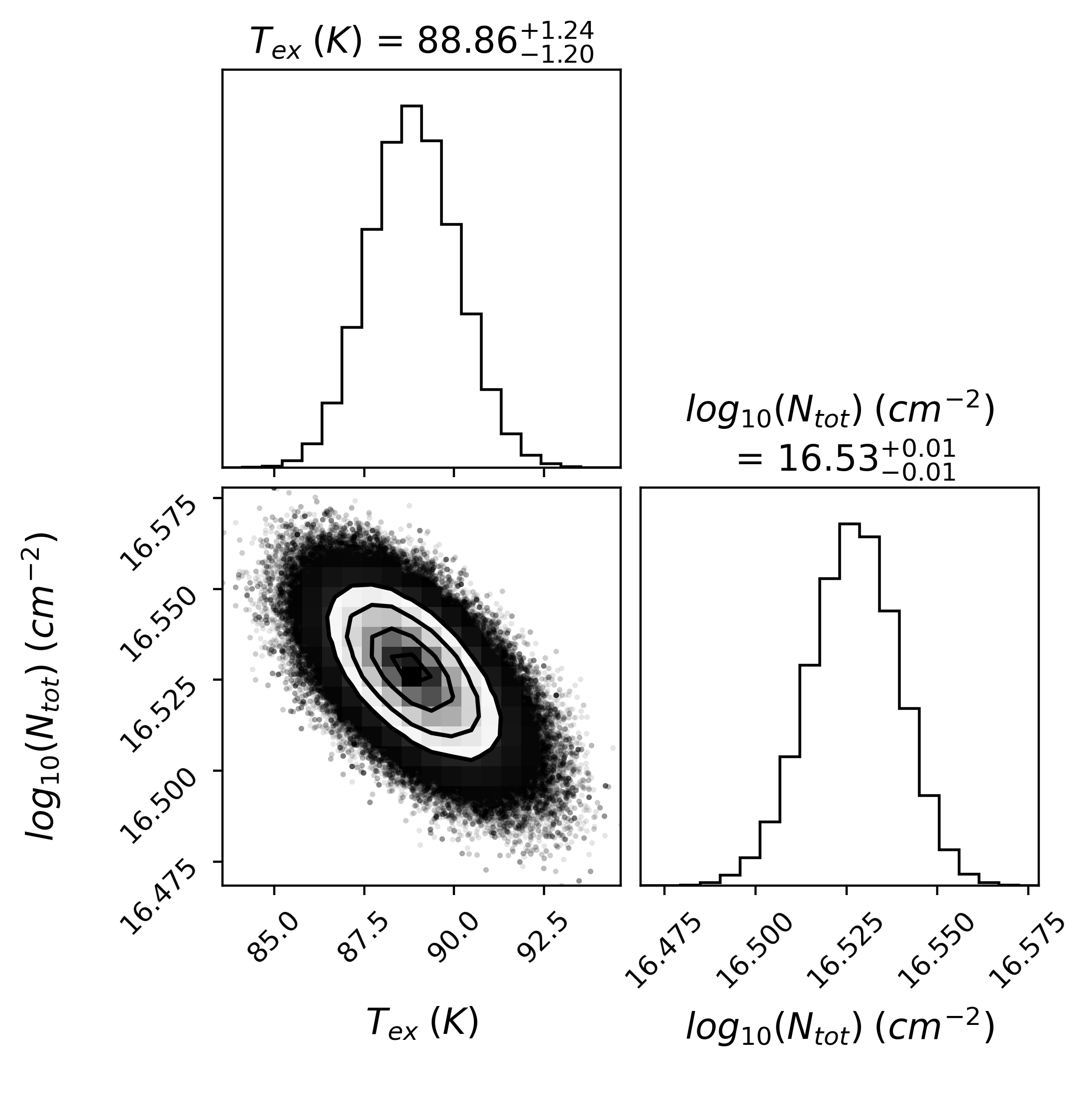}
\caption{ Glycolonitrile $\mathrm{HOCH_2CN}$ \label{HOCH2CN}}
\end{figure}

\clearpage
\section{Appendix C}\label{AppendixC}
\begin{figure}[!htbp]
\includegraphics[width=1.0\textwidth]{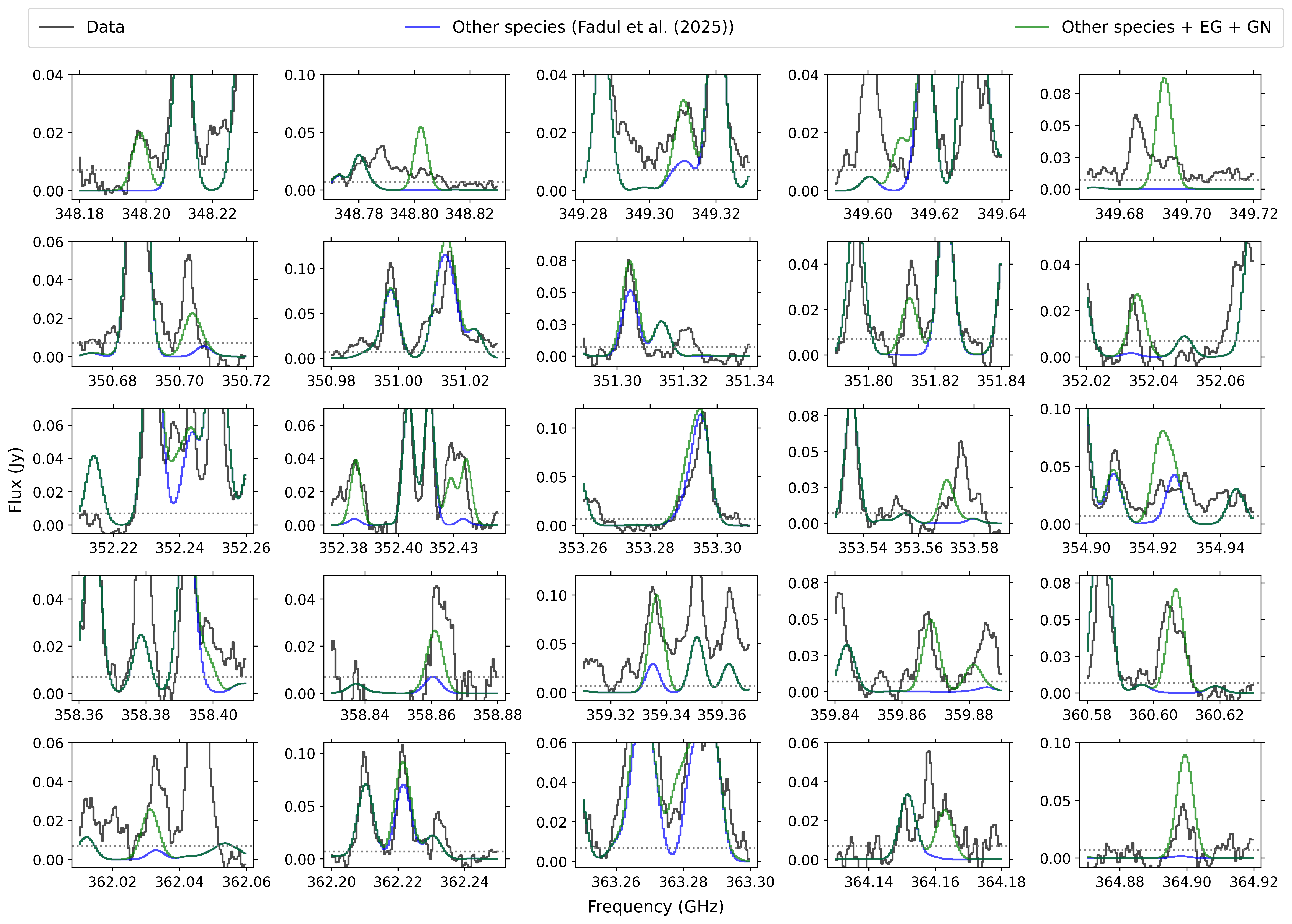}
\caption{Represent the cumulative model for all species reported by \cite{Fadul25} in blue, and the same cumulative model include EG and GN in green. The black represents the observed spectra, and the gray dotted horizontal line represents the noise level, which corresponds to 1$\sigma$. \label{summed_spectra}}
\end{figure}

Figure~\ref{summed_spectra} shows the cumulative model, which adds the best-fit spectra from each molecule reported by \cite{Fadul25} in blue. Additionally, the EG and GN best-fit spectra are included in the same cumulative model, shown in green.  

\clearpage
\section{Appendix D}\label{AppendixD}

Here we show the predicted spectra from our model (see \cite{Fadul25} for more details) of ethylene glycol ($\mathrm{g'Gg-(CH_2OH)_2}$) at $\mathrm{T_{ex}}$ of 300 K and $\mathrm{N_{tot}}$ of $3.6\times10^{16}\; \mathrm{cm^{-2}}$ (best fit values of EG from Band 7 observations (this work)) plotted on top of ALMA Band 6 data published by \cite{Lee2025ApJ}. The data have been downloaded from the ALMA archive, and the spectrum was extracted using the function \textit{integrated\_spectrum()} implemented in the Python package \textit{Gofish} \citep{GoFish} (for more details see \cite{Fadul25} Section 3). The main parameters used to extract the spectrum are an inner radius of $0\arcsec$, an outer radius of $0.21\arcsec$, a position angle of $32^\circ$, an inclination angle of $38.3^\circ$, a distance of 388 pc, and a stellar mass of 1.29 $\mathrm{M\odot}$ (\cite{Cieza2016Natur, Lee_2019, Tobin23, Yamato2024AJ}).

\begin{figure}[!htbp]
\includegraphics[width=0.85\textwidth]{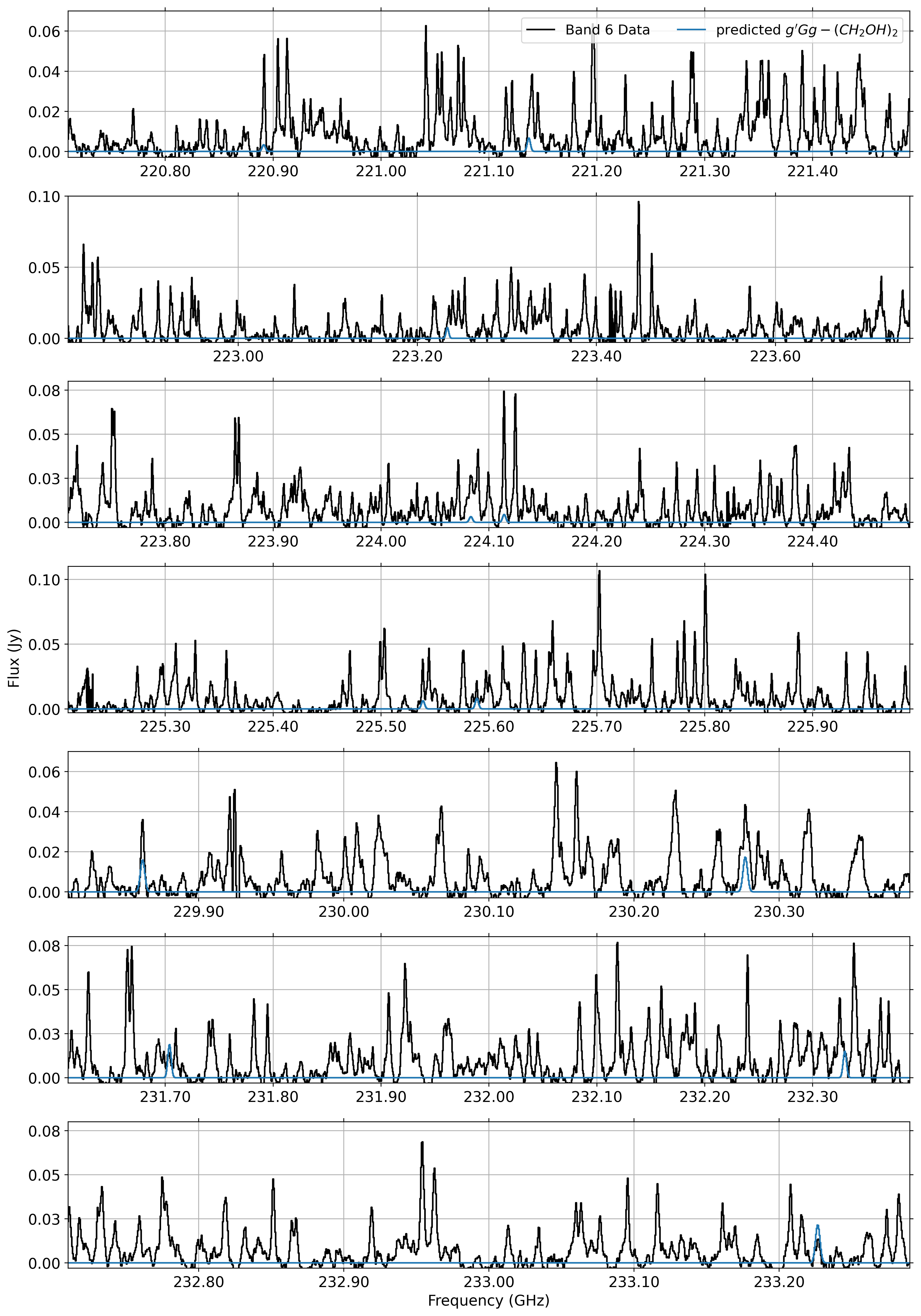}
\caption{Shows ethylene glycol spectra in steel blue overlaid on the observed spectra from ALMA Band 6 in black  \citep{Lee2025ApJ}.\label{firstFigure}}
\end{figure}

\begin{figure}[!htbp]
\includegraphics[width=0.85\textwidth]{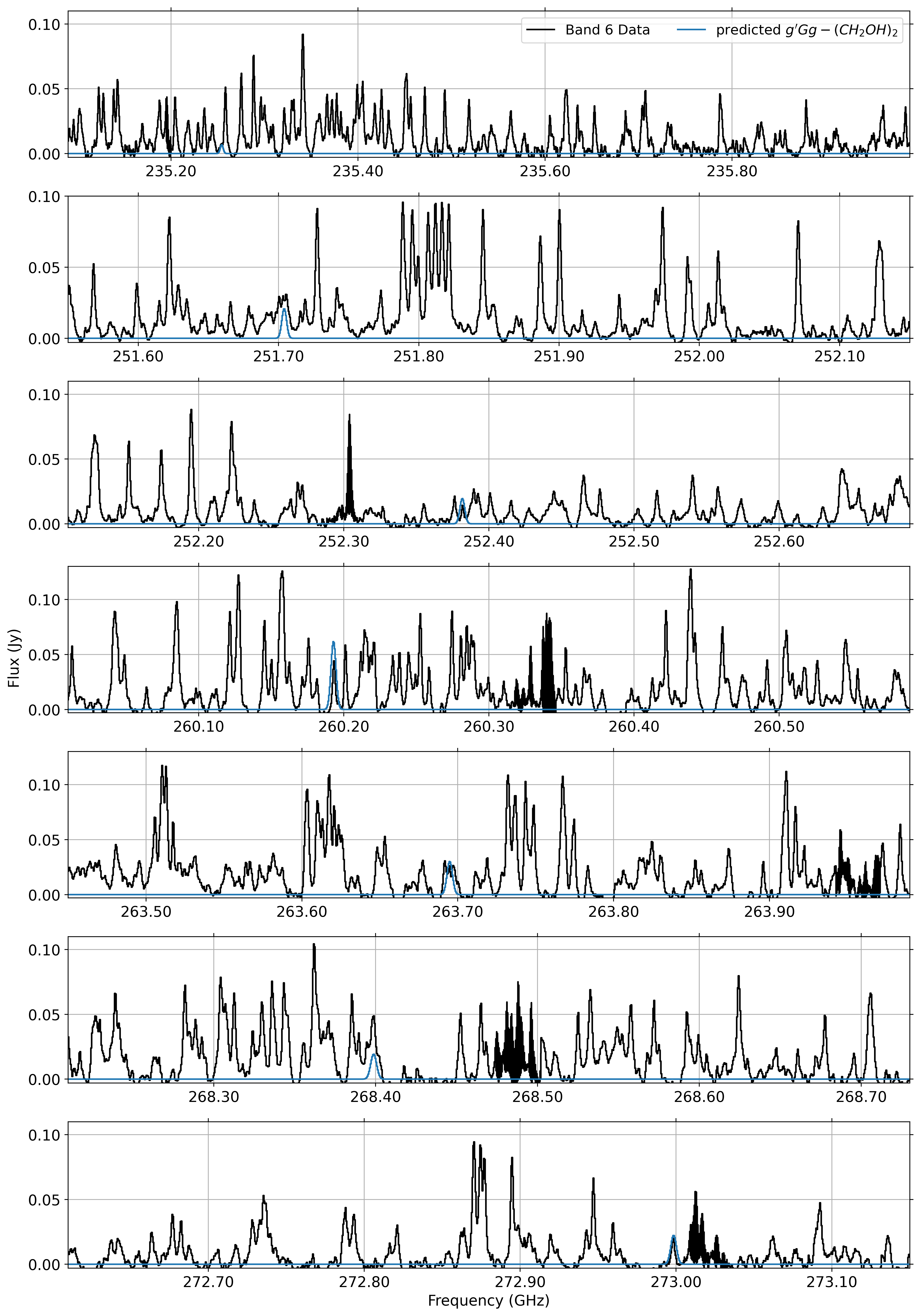}
\caption{Continuation of Figure~\ref{firstFigure}}
\end{figure}

\end{document}